# Role of Chalcogen atoms in In Situ Exfoliation for Large-Area 2D Semiconducting Transition Metal Dichalcogenides

Zhiying Dan[1], Ronak Sarmasti Emami[1], Giovanna Feraco[1], Melina Vavali[1], Dominic Gerlach[1], Petra Rudolf[1] and Antonija Grubišić-Čabo[1,*]

[1] *Zernike Institute for Advanced Materials, University of Groningen, 9747 AG Groningen, The Netherlands.*

Correspondence*:
Corresponding Author
a.grubisic-cabo@rug.nl

**ABSTRACT**

Two-dimensional (2D) transition metal dichalcogenides have emerged as a promising platform for next-generation optoelectronic and spintronic devices. Mechanical exfoliation using adhesive tape remains the dominant method for preparing 2D materials of highest quality, including transition metal dichalcogenides, but always results in small-sized flakes. This limitation poses a significant challenge for investigations and applications where large scale flakes are needed. To overcome these constraints, we explored the preparation of 2D $WS_2$ and $WSe_2$ using a recently developed kinetic *in situ* single-layer synthesis method (KISS). In particular, we focused on the influence of different substrates, Au and Ag, and chalcogen atoms, S and Se, on the yield and quality of the 2D films. The crystallinity and spatial morphology of the 2D films were characterized using optical microscopy and atomic force microscopy, providing a comprehensive assessment of exfoliation quality. Low-energy electron diffraction verified that there is no preferential orientation between the 2D film and the substrate, while optical microscopy revealed that $WSe_2$ consistently outperformed $WS_2$ in producing large monolayers, regardless of the substrate used. Finally, X-ray diffraction and X-ray photoelectron spectroscopy demonstrate that no covalent bonds are formed between the 2D material and the underlying substrate. These results identify KISS method as a non-destructive approach for a more scalable approach of high-quality 2D transition metal dichalcogenides.

## 1 INTRODUCTION

In recent years, two-dimensional (2D) materials have gained significant attention due to their unique electronic and optical properties, such as the transition from an indirect to a direct band gap when going from bulk to a single layer of $MoS_2$, which makes them highly promising for electronic, optoelectronic and spintronic devices (1; 2; 3; 4; 5). Despite their potential, the efficient and scalable production of high-quality monolayers remains a significant challenge. Mechanical exfoliation (ME), which uses adhesive tape, is widely recognized for producing high-quality flakes; however, the size of these flakes is usually on the order of tens of micrometers, making them inadequate for many surface science studies and practical applications. Chemical vapor deposition (CVD), on the other hand, can be used to synthesize wafer-sized monolayers (6), but it involves complex optimization processes, requires high temperature conditions, and often results in monolayers with high defect density and lower quality compared to ME. Additionally, CVD lacks the universality of ME in its applicability to various 2D materials. Molecular beam epitaxy (MBE) also suffers from similar drawbacks (7), including the need for stringent optimization of growth parameters and challenges in achieving defect-free monolayers. As a result of this, ME remains the preferred method for the preparation of 2D materials, particularly in a research setting, due to its simplicity, efficiency, low cost and ability to produce monolayers of the highest quality (8). However, in addition to the heterogeneous interlayer forces and surface interactions that often restrict flake sizes to tens of micrometers, ME also suffers from poor yield (9), resulting in a dramatic variations in the thickness and quality of exfoliated layers over the whole substrate touched by the adhesive tape.

The challenges are exacerbated when handling air- or moisture-sensitive 2D materials, such as black phosphorus and $VSe_2$, which degrade rapidly when exposed to ambient conditions (10; 11). To address these issues, ambient-sensitive materials must be handled in controlled environments, such as glove boxes, further complicating the exfoliation process (12). Transferring sensitive materials from glove boxes to a measurement system without environmental control is another major obstacle, requiring a sealed transfer system or the use of encapsulation techniques (13), which adds complexity and risk of contamination or damage.

Recently, the metal-assisted mechanical exfoliation technique has emerged as a feasible solution for large-area 2D material production (14; 15; 16; 17). By exploiting the stronger adhesion between the metals and the outermost layer of bulk layered crystals, this technique can produce monolayers on a millimeter–or even centimeter–scale by pressing a tape containing a bulk crystal onto clean, smooth metal substrates, such as gold (17). However, this method is typically performed under ambient conditions, making it unsuitable for air-sensitive materials. Moreover, this method is incompatible with surface techniques that require *in situ* surface cleaning.

Here, we employed an ultra-high vacuum (UHV) exfoliation method (16) to prepare large-area monolayers of transition metal dichalcogenide (TMDC) films. Using this approach we successfully produced near-millimeter-scale monolayers of $WS_2$ and $WSe_2$ on metal substrates (Au, Ag). Low-energy electron diffraction (LEED) was used to explore existence of potential preferred angle between the layered film and the metal substrates, while optical microscopy was used to study the exfoliation yield of monolayer $WSe_2$ than $WS_2$. Atomic force microscopy (AFM) was employed to verify the thickness of exfoliated flakes. Finally, X-ray photoelectron spectroscopy (XPS) and X-ray diffraction were used to investigate chemical bonding and substrate quality following UHV exfoliation. Our findings demonstrate the feasibility of the UHV exfoliation method for producing high-quality, large-area monolayer TMDCs, as the nearly millimeter-scale exfoliated flakes retain their crystallinity and are well-suited for *in situ* surface studies. Importantly, the size of the flakes correlates with the quality of the parent crystals, highlighting the potential of this technique for advancing scalable, high-quality 2D material production.

## 2 MATERIALS AND METHODS

### 2.1 Materials

$WS_2$ and $WSe_2$ bulk crystals were purchased from HQ Graphene, with a lateral size of ≈5 mm. The bulk crystals were cut to suitable dimensions to fit customized holders and mounted using vacuum-compatible silver epoxy (EPO-TEK). Prior to loading into the load-lock chamber, the surface of each crystal was tape-cleaved to ensure flatness of the surface. Following this, a pre-prepared adhesive tape (Lyreco,

INVISIBLE TAPE) was attached to the top of the crystal for subsequent mechanical cleavage under UHV conditions. Ag(111)/mica and Au(111)/mica (Georg-Albert-PVD) substrates were cut to the appropriate size and fixed to a flag style sample plate by welded tantalum strips. The Au and Ag substrates were cleaned using repeated cycles of Ar$^+$ sputtering (45-min at $1.0 \times 10^{-6}$ mbar and 1.5 kV) and annealing (30-min at 600 K).

## 2.2 Characterization Techniques

### 2.2.1 Low-Energy Electron Diffraction

LEED measurements (SPECS ErLEED 1000A) with a 1 mm spot size were performed at room temperature under a pressure of $\approx 5 \times 10^{-10}$ mbar, using electron energy of 125 eV for all samples. LEED was used to verify the cleanliness of the Au(111) and Ag(111) substrates following the cleaning procedure (prior to exfoliation) and to confirm the success of exfoliation after the exfoliation process.

### 2.2.2 Optical microscopy

Monolayers of exfoliated materials were identified using optical contrast technique. Optical images were acquired with an Olympus microscope and processed using ImageJ software to analyze regions corresponding to monolayer contrast. The images were split into red, green and blue (RGB) channels, with the red channel selected for analysis due to its excellent contrast. ImageJ threshold routines were applied to identify flakes and calculate monolayer areas as well as lateral sizes, represented by the maximum Feret diameter. Statistical analysis of the size distributions was performed by fitting the data to a log-normal distribution.

### 2.2.3 Atomic Force Microscopy

AFM was conducted under ambient conditions with Dimension FastScan Bruker and Cypher S AFM (Asylum Research) microscopes using the AC160-TSA tip. The measurements were performed in tapping mode. The AFM was calibrated using a standard calibration sample prior to the experiments. AFM data analysis was carried out using the Gwyddion software package.

### 2.2.4 X-ray photoelectron spectroscopy

XPS measurements of the TMDC/metal samples were performed using an SSX-100 (Surface Science Instruments) spectrometer equipped with a monochromatic Al K$\alpha$ X-ray source (hv= 1486.6 eV). The measurement chamber pressure was maintained at $1 \times 10^{-9}$ mbar during data acquisition, with a photoelectron take-off angle of 37 ° relative to the surface normal. The analyzed area had a diameter of 1 mm$^2$ and the energy resolution was 1.6 eV. The binding energy (BE) values were reported with an accuracy of 0.1 eV and referenced to the C 1$s$ core level of adventitious carbon at 284.91 eV (18). All XPS spectra were analyzed using the least-squares curve-fitting program Winspec (developed at LISE, University of Namur, Belgium). Spectral fitting included Shirley (for W 4$f$) and liner background subtraction (for S 2$p$ & Se 3$p$). Because the S 2$p$ and Se 3$p$ peaks lie on the intense Au W 4$f$ background, a linear background subtraction was deemed sufficient for accurately isolating the peak signals(19). The peak profiles were modeled as a convolution of Gaussian and Lorentzian functions using the minimum number of peaks necessary to reflect the chemical structure of the sample, while accounting for experimental resolution. The uncertainty in the peak intensity determination was within 2 % for all of the core-level spectra, with measurements taken at two points for each sample.

# 3 RESULTS AND DISCUSSION

## 3.1 Exfoliation of monolayers in UHV

Samples were prepared using the kinetic *in situ* single-layer synthesis (KISS) method, as illustrated in **Figure 1** for the case of WSe$_2$ on Ag(111). The KISS exfoliation process, along with the pre-treatment of bulk crystals and metallic substrates (detailed in the Materials and Methods section), was conducted under UHV conditions maintaining a pressure better than $1 \times 10^{-9}$ mbar. **Figure 1(a)** depicts preparation of a clean bulk surface of WSe$_2$ by adhesive tape cleaving in UHV. To ensure that the surface is as clean as possible, this step is performed shortly before the KISS exfoliation takes place. **Figure 1(b)** illustrates the substrate preparation process, which involved multiple cycles of Ar$^+$ sputtering and annealing to remove surface impurities and moisture (20). After three cleaning cycles, the substrate was naturally cooled to room temperature. At this point, KISS exfoliation was performed by bringing a freshly cleaved WSe$_2$ bulk crystal into contact with the substrate, **Figure 1(c)**, followed by the controlled separation of bulk crystal from the substrate, **Figure 1(d)**. Due to the strong adhesion between the metallic substrate and TMDC (14), in this case WSe$_2$, the topmost layer–or, in some cases, a few layers–is exfoliated onto the substrate, **Figure 1(e)**. An example of KISS-exfoliated monolayer (ML) WSe$_2$ on Ag(111) is shown in **Figure 1(f)**, with the ML flake (outlined in green) having a lateral size of approximately 500 $\mu$m. Typically, when the crystal was significantly smaller than the substrate, exfoliation was performed at several distinct locations on the substrate surface to increase the coverage.

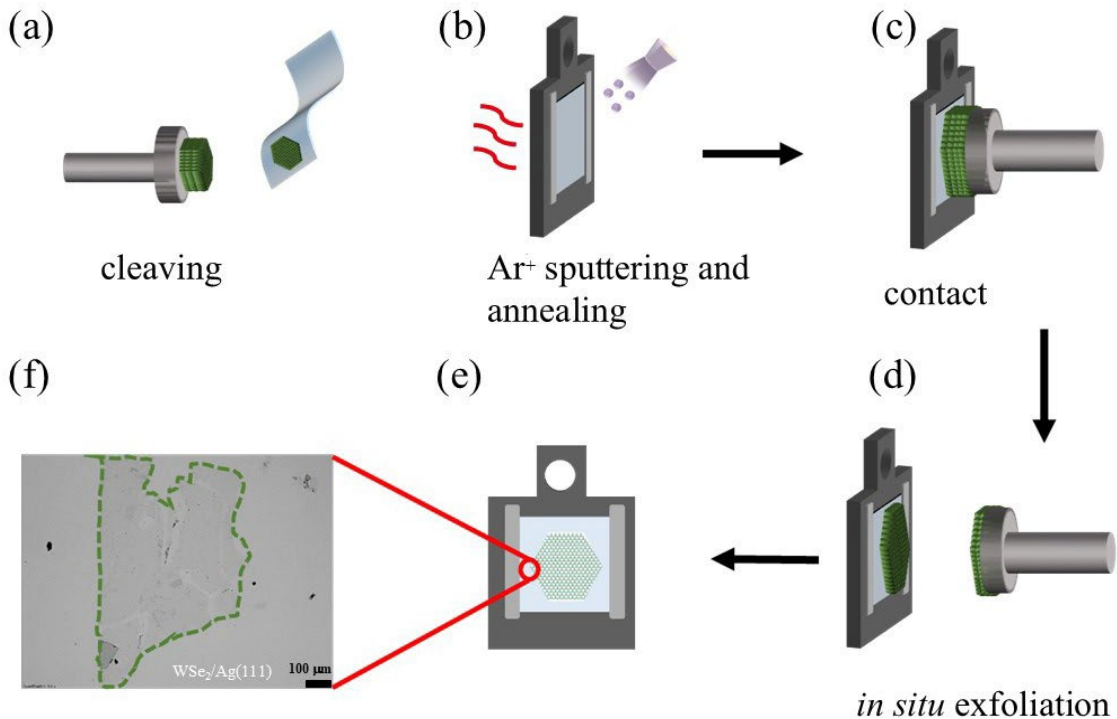

**Figure 1.** A schematic of the kinetic *in situ* single layer synthesis (KISS) exfoliation process: (a) preparation of an ultra-flat, clean WSe$_2$ surface. (b) sputtering and annealing of the Ag(111) substrate to achieve an atomically clean and flat surface. Freshly prepared surfaces are brought into contact (c), followed by careful separation (d) to enable *insitu* exfoliation of the 2D material onto the substrate (e). This process results in deposition of a high-quality, large-area monolayer 2D material, verified by optical microscopy, with single-layer WSe$_2$ successfully transferred onto the Ag(111) substrate (f).

## 3.2 Surface Morphology and Structure

### 3.2.1 Angular alignment with the substrate

Following exfoliation procedure, *in situ* characterization was conducted using the LEED system mounted on the same UHV chamber. As shown in **Figure 2**, LEED patterns were obtained for four individual flakes exfoliated onto Au(111), **Figure 2(a)**, and Ag(111) substrates, **Figure 2(b)**. All pattern exhibited distinct nested hexagonal diffraction patterns, consistent with the hexagonal crystalline lattices of $WSe_2$, Ag(111) and Au(111). The outer hexagons in **Figure 2** originate from the substrates (Au or Ag), while the inner hexagons come from $WSe_2$. The presence of both hexagonal patterns confirms the successful exfoliation of $WSe_2$ (16). Notably, the sharper and brighter diffraction spots observed in $WS_2$ compared to $WSe_2$ can be attributed to its higher lattice phonon frequency (21), which inherently corresponds to smaller amplitudes of atomic thermal vibrations and results in reduced attenuation of elastic scattering intensity (22).

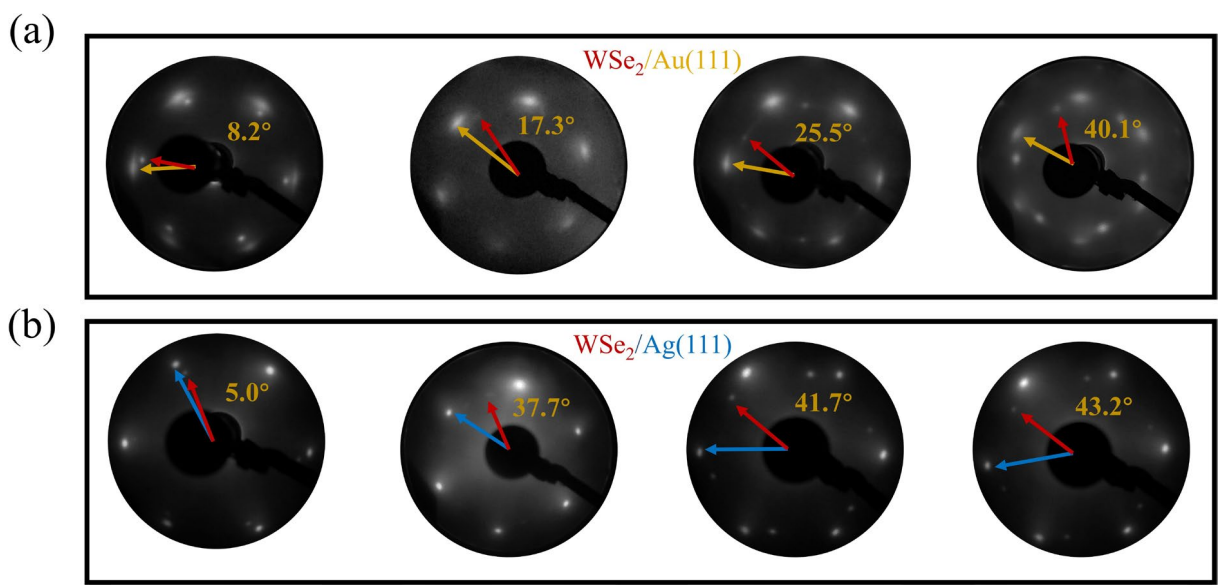

**Figure 2.** Low-energy electron diffraction (LEED) analysis of the rotational alignment between the exfoliated 2D material and the substrate: (a) LEED images of $WSe_2$ on Au(111); (b) LEED images of $WSe_2$ on Ag(111). All images were obtained with an electron energy of 125 eV. The red arrow points to the diffraction pattern of $WSe_2$, and the yellow and blue arrows indicate the diffraction features of Au and Ag, respectively. No preferred rotational alignment is observed on either substrate.

Incidentally, the diffraction spots from the Au(111) substrate are significantly broader than those from the Ag(111) substrate. This broadening, also observed for $WS_2$/Au(111) samples (see Supplementary **Figure S1(a)**), suggests that the coherently diffracting domains are smaller in the case of Au(111) thin films grown on mica than in the Ag(111) ones. Moreover, one also notices that the Au(111) spots are slightly elongated, contrary to the Ag(111) spots which are round. This points to a slight misalignment between Au(111) grains and higher step density (23; 24).

Epitaxially grown 2D materials typically exhibit preferred stacking orientation relative to the substrate they are grown on (25). However, KISS exfoliation does not favor a specific stacking orientation alignment between the exfoliated 2D materials and its substrates, as seen from the LEED patterns. **Table 1** summarizes the twist angles derived from the **Figure 2** and **Supplementary Figure S1**, showing significant angular variations between the TMDC flakes and the substrates, with angles randomly distributed between 0° and 60°. Furthermore, lattice matching between the metal substrate and the TMDC seems to have no effect on the exfoliation process. In other words, we can exclude that grains in the parent crystal of Au (Ag) with a specific orientation with respect to the TMDC substrate attach more easily (24; 26).

|  | Sample 1 | Sample 2 | Sample 3 | Sample 4 |
|---|---|---|---|---|
| $WS_2$/Ag(111) | 4.4° | 9.2° | 28.1° | 34.4° |
| $WS_2$/Au(111) | 8.7° | 20.2° | 47.1° | 56.8° |
| $WSe_2$/Ag(111) | 5.0° | 37.7° | 41.7° | 43.2° |
| $WSe_2$/Au(111) | 8.2° | 17.3° | 25.5° | 40.1° |

**Table 1.** Summary of twist angles between TMDCs and substrates based on LEED images of $WS_2$/Au, $WS_2$/Ag, $WSe_2$/Au, and $WSe_2$/Ag. In each case four individual samples were examined. No preferred orientation was observed for any of the samples.

3.2.2    Role of chalcogen atom

The presence of different chalcogen atoms influences various intrinsic properties of TMDCs (8), but their influence on KISS exfoliation trends remains unclear. **Figure 3** presents the exfoliation results of $WS_2$ and $WSe_2$ on Ag(111) substrates, as observed by optical microscopy. Despite using the same type of substrate and proceeding with the exfoliation in an identical manner, differences in exfoliation yield between the two materials are clearly evident. For $WSe_2$/Ag(111) sample, shown in **Figure 3(a)**, large and crack-free monolayer flakes were consistently produced, up to a maximum lateral size of 750 $\mu$m. Only a few flakes exhibited thicknesses exceeding that of a monolayer, and in some instances, minor bulk structures were observed at the edges of flakes. In contrast, $WS_2$/Ag(111), shown in **Figure 3(b)**, displayed a more fragmented morphology with smaller flake size and a significant presence of few-layer and bulk structures. This fragmentation and presence of thicker structures makes it challenging to isolate individual monolayer flakes suitable for further characterization. To rule out substrate effects, the same protocol was applied to $WS_2$/Au(111) and $WSe_2$/Au(111) samples (shown in Supplementary **Figure S2**). The results confirmed that $WS_2$ consistently exhibited greater fragmentation, further emphasizing the role of the chalcogen elements in facilitating exfoliation.

To quantitatively evaluate the yield of monolayer flakes, we measured and analyzed the size distribution of flakes from four different samples. Given the predominantly irregular morphology of the exfoliated flakes, the Feret diameter was used to determine their lateral dimensions (15; 27). Using ImageJ software, the Feret diameters were extracted by manually applying an appropriate threshold, which meant setting the contrast range manually, where pixels above the threshold are considered as TMDC, while those below are attributed to the substrate (28; 29). As shown in **Figure 3(c)**, the statistical size distribution for the four $WSe_2$/Ag(111) samples reveals that most flakes fall within the 200-250 $\mu$m range, with the largest lateral dimension reaching up to 750 $\mu$m. In contrast, the statistical analysis for the four $WS_2$/Ag(111) samples, **Figure 3(d)**, indicates that the flakes are primarily within the 100-150 $\mu$m range, with a maximum size of 350 $\mu$m, and display a higher degree of fragmentation. The same analysis for $WSe_2$ and $WS_2$ on Au(111) can be found in the Supplementary information, **Figures S2(c),(f)** and is in agreement with the results on Ag(111).

The larger dimensions of WSe$_2$ flakes, the absence of cracks in the transferred WSe$_2$ flakes and higher overall coverage of the substrate indicate a more stable and robust interaction with the noble metal during the KISS process. This behavior can be attributed to the larger atomic radius and higher polarization of Se atoms, which result in stronger van der Waals interaction between the WSe$_2$ flakes and the substrate (8). Additionally, the lower electronegativity of Se facilitates the charge transfer between WSe$_2$ and the substrate, thereby enhancing the charge coupling effect at the interface (30). It is worth noting that these outcomes may also be influenced by the quality of the parent crystals, as the bulk WSe$_2$ crystal appeared to have higher quality with fewer bulk defects, **Figure S6**.

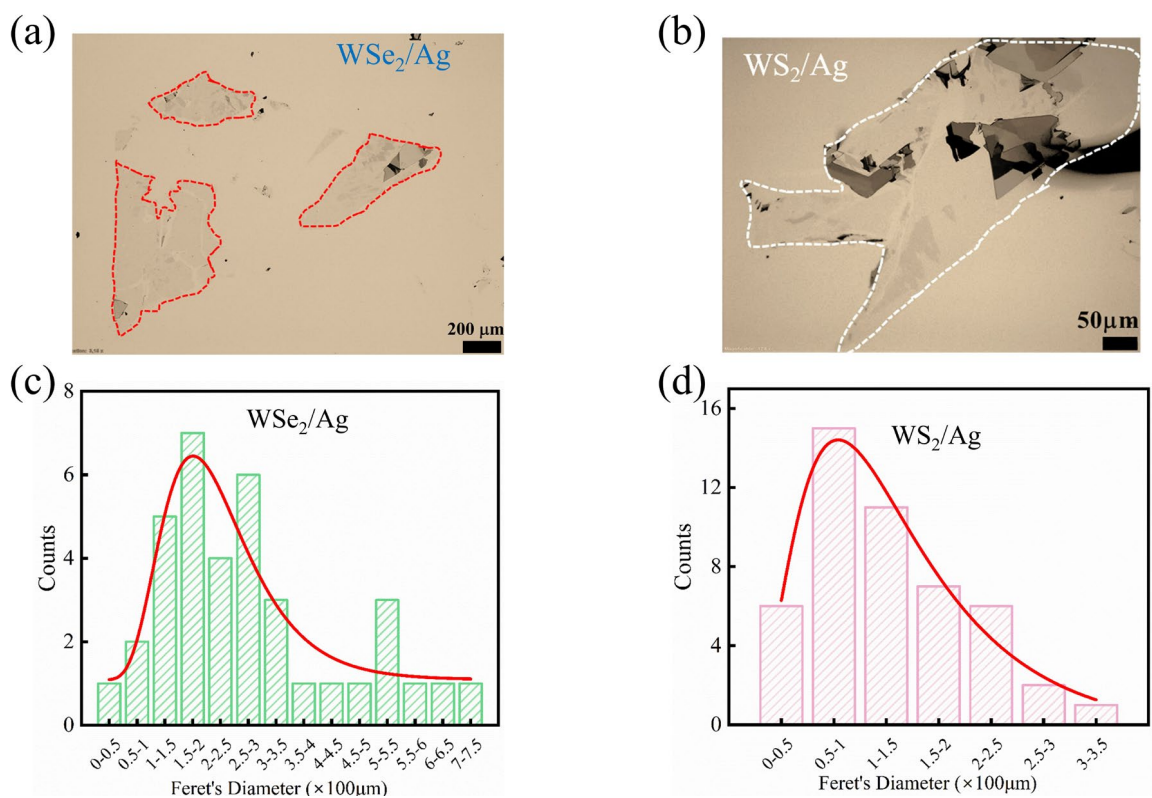

**Figure 3.** Optical microscopy characterization of exfoliated flake size distribution: (a) and (b) are representative optical images of WSe$_2$/Ag(111) and WS$_2$/Ag(111), respectively. (c) Distribution of flake lateral size ( maximum Feret diameter) of the optical image for four WSe$_2$/Ag(111) samples. Optical images of the other samples are shown in **Figure S3**. (d) Distribution of flake lateral size for four WS$_2$/Ag(111) samples. Optical images of the other samples are shown in **Figure S3**. Histograms showing the counts of different lateral sizes of flakes are fitted using a log-normal distribution, visualized by the solid red line. The p-values are all less than 0.05, indicating that the log-normal distribution is suitable.

3.2.3   Surface morphology

To investigate the flake thickness, optical microscopy, **Figure 4(a)**, and AFM, **Figure 4(b)**, were used. The distinct color distribution observed in the optical images clearly indicates regions of varying thickness (29). The step profile analysis showed that the monolayer thickness is approximately 0.82 nm, which is slightly larger than previously reported values for monolayer WSe2 (16). This difference in height can be attributed to AFM instrumental offset, as well as the presence of adventitious carbon, impurities or moisture on the TMDC surface from exposure to air before and during AFM measurements (31; 32; 33). By combining AFM measurements and optical microscopy data, the color of the flake can be correlated with its thickness, as shown in **Figure 4(a)**, allowing for faster identification of the layer number, similar to what has been done for mechanically exfoliated 2D materials on SiO2/Si (29).

In the AFM topography, a distinct regular pattern can also be observed, marked with red arrows in **Figure 4(b)**, which appears to originate from the underlying substrate. The angle between the lines in the pattern is exactly 60°, and the pattern arises from the local reconstruction of surface steps along the symmetric crystallographic directions of the Ag(111) surface after annealing, resulting in the formation of periodic stripes (34).

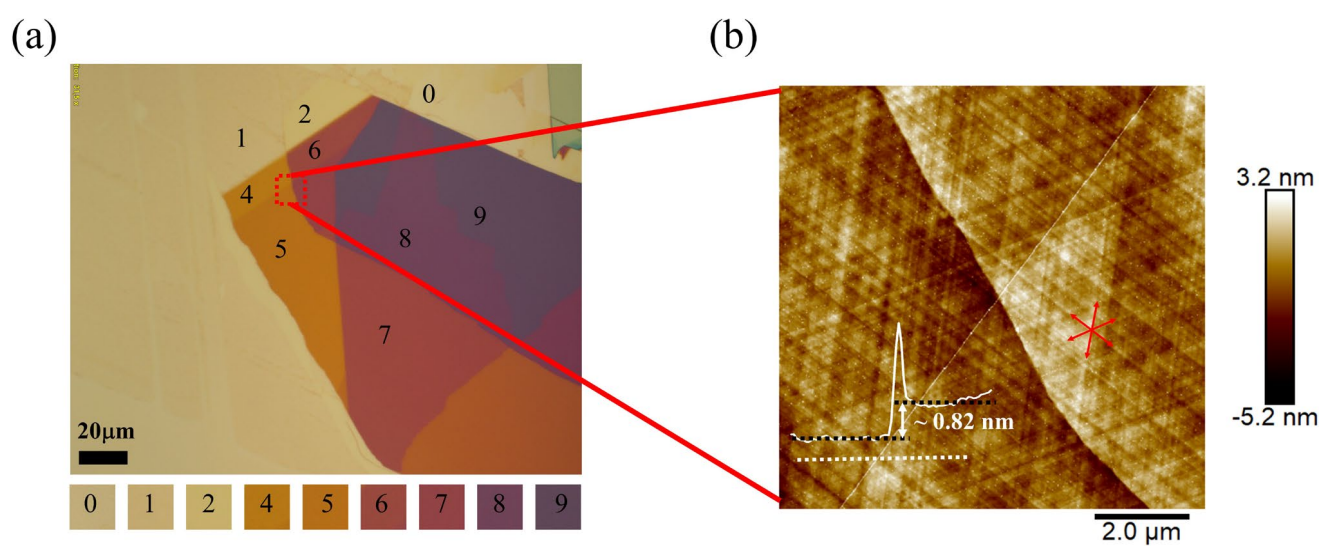

**Figure 4.** Thickness characterization based on color variations observed in the optical microscopy image: (a) optical image of the WSe2/Ag(111). Different layers correspond to different colors and are indicated by numbers 0–9, where 0 represents bare Ag and 9 corresponds to nine layers of WSe2. (b) AFM measurements conducted within the red square outlined in (a). The white solid line represents the step height (≈ 0.82 nm) along the dashed line, confirming a single-layer thickness difference between adjacent layers.

## 3.3 Chemical Composition and Interfacial Property

We performed XPS measurements to better understand the interaction between TMDCs and metal surfaces. WSe2 and WS2 were measured on Ag(111) and Au(111) substrates, respectively. **Figure 5(a)** presents the wide scan spectrum of WSe2/Au(111), which is used to identify the elements present on the surface. The characteristic peaks corresponding to the W $4f$, W $4d$, Se $3d$, and Se $3p$ orbitals confirm the presence of WSe2, and demonstrate that KISS exfoliated flakes are large enough to be located and measured with a 1 mm beam. The C $1s$ peak indicates the presence of adventitious carbon, likely resulting from the transfer in air from the exfoliation chamber to the XPS chamber (15).

**Figures 5(b) and 5(c)** show detailed spectra for W $4f$ and Se $3p$. The asymmetry observed in the W $4f$ peak is attributed to shake-up events, promoting conduction electrons from below to above the Fermi level. The Se $3p_{1/2}$ and Se $3p_{3/2}$ lines are peaked at binding energies of 161.21 eV and 167.21 eV, respectively.

The W $4f_{7/2}$ and W $4f_{5/2}$ lines are peaked at binding energies of 31.6 eV and 33.8 eV, respectively. The binding energy of W $4f$ for WSe$_2$ and WS$_2$ on Au and Ag substrates as summarized in **Table S1**, shows slight variations in different samples, remains consistent with previous reports (35; 36; 37; 38). Since no components related to WO$_x$ can be seen in the W $4f$ spectra, nor components relative to oxidized selenium (162 eV (39) ) and sulfur(168.8 eV (40) ) are found, we conclude that KISS exfoliated WSe$_2$ and WS$_2$ layers are air-stable, similar to MBE grown samples (41; 42). Although changes in XPS binding energy suggest some degree of charge transfer, no clear evidence of Se (S)-Au (Ag) covalent bonding was observed in the XPS spectra. Similar observations have been reported for MBE grown samples (43), indicating a weaker van der Waals type bonding.

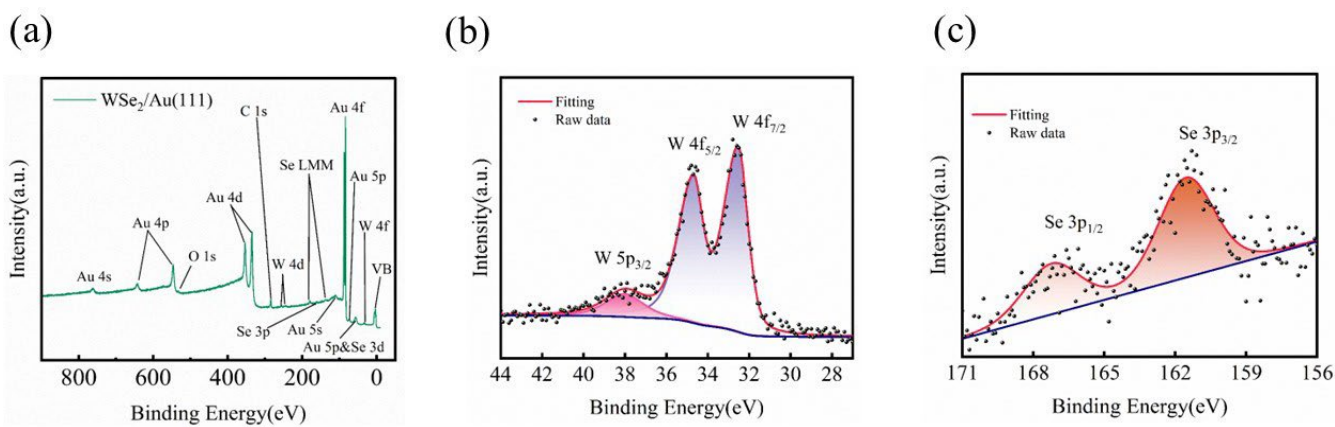

**Figure 5.** X-ray photoelectron spectroscopy (XPS) characterization of WSe$_2$/Au(111): (a) Wide scan of WSe$_2$/Au(111) showing the expected Au, W and Se peaks. The presence of C and O peaks suggests potential contamination from adsorption during the transfer process which was performed in air, or insufficient annealing prior to measurement. Detailed scans of the (b) W $4f$ and (c) Se $3p$ core level regions (dots) with their corresponding fits (continuous line). No signs of oxidation are observed for either W or Se. The XPS spectra of W were background-subtracted using the Shirley background, while from the Se spectra a linear background was subtracted because it sits on the intense Au $4f$ background.

Since KISS exfoliation involves the mechanical contact of a bulk crystal with another solid surface, it is essential to assess whether the process induces any damage to the substrate. To evaluate the impact of exfoliation on the substrate, we measured the (111) Bragg reflection of a single-crystal silver substrate in a near-backscattering geometry (see Section 4 in SI), which is sensitive to deformation of the substrate lattice. The measurement was performed with a focused beam aligned to a WSe$_2$ flake after KISS procedures. The width of the Ag(111) reflectivity curve, shown in **Figure S5 (d)**, was determined to be 0.95 eV, which is identical to the intrinsic width of the Ag(111) reflection convoluted with the energy width of the optics, indicating that the exfoliation process of WSe$_2$ has a negligible effect on the crystalline of the substrate. This finding underscores the suitability of KISS exfoliation for preparing high-quality monolayer TMDCs while preserving the substrate's properties for subsequent characterization and applications.

## 4 CONCLUSIONS

The KISS exfoliation method has proven to be a reliable method for producing high-quality large-area monolayers of WS$_2$ and WSe$_2$, on metallic substrates in ultra-high vacuum. Notably, exfoliation of WSe$_2$ yielded crack-free flakes with larger lateral dimensions and a higher overall coverage of the substrate than the exfoliation of WS$_2$, presumably due to the higher quality of WSe$_2$ bulk crystal and to selenium's advantageous atomic properties, including its larger radius, higher polarization, and lower electronegativity. By combining optical microscopy and atomic force microscopy, we demonstrate that optical microscopy is a practical method for layer identification based on color contrast. Structural analysis using low-energy electron diffraction revealed that there is no preferred angle between the substrates and the two-dimensional transition metal dichalcogenides, implying that lattice matching between the substrate and transition metal dichalcogenide seems to have no effect on the exfoliation process, and thus one can exclude that grains in the parent crystal with a specific orientation with respect to the substrate attach more easily. Chemical analysis by X-ray photoelectron spectroscopy verified the air-stability of the exfoliated films, and no signs of covalent bonding between the substrate and transition metal dichalcogenides were found, with only mild charge transfer effects observed. This suggests that while interaction between the substrate and the topmost transition metal dichalcogenide layer is stronger than the transition metal dichalcogenide interlayer bonds, it is not of covalent nature. Furthermore, X-ray diffraction confirmed that the exfoliation process has no impact on the crystalline quality of the metallic substrate, demonstrating the method's non-destructive nature. These results validate kinetic *in situ* single-layer synthesis method as an excellent technique for fabricating high-quality, large size crack-free monolayer transition metal dichalcogenides on metallic substrates without substrate degradation.

## CONFLICT OF INTEREST STATEMENT

The authors declare that the research was conducted in the absence of any commercial or financial relationships that could be construed as a potential conflict of interest.

## ACKNOWLEDGMENTS

AGC acknowledges the financial support of the Zernike Institute for Advanced Materials and the research program "Materials for the Quantum Age" (QuMat) for financial support. This program (registration number 024.005.006) is part of the Gravitation program financed by the Dutch Ministry of Education, Culture and Science (OCW). ZD acknowledges the fellowship from the Chinese Scholarship Council (No.202206750016). The authors would like to thank Diamond Light Source for beamtime (proposal SI35796) and the staff of beamline I09, Deepnarayan Biswas and Tien-Lin Lee, for their assistance with experiments and data collection.

# *Supplementary Material*

## 1 LOW-ENERGY ELECTRON DIFFRACTION

**Figure S1** shows the LEED patterns of $WS_2$ exfoliated onto Au(111) and Ag(111) by the KISS method in UHV. The Au(111) diffraction spots appear blurred and elongated compared to the Ag(111) spots due to a higher degree of the smaller grains, higher step density, and a higher degree of polycrystallinity of the Au(111) thin films, as discussed in the main text (1; 2). The two hexagonal patterns, rotated at different angles with respect to each other, are clearly visible, illustrating that $WS_2$ was successfully exfoliated using the KISS method. The angle between the $WS_2$ and the substrate is indicated for each flake.

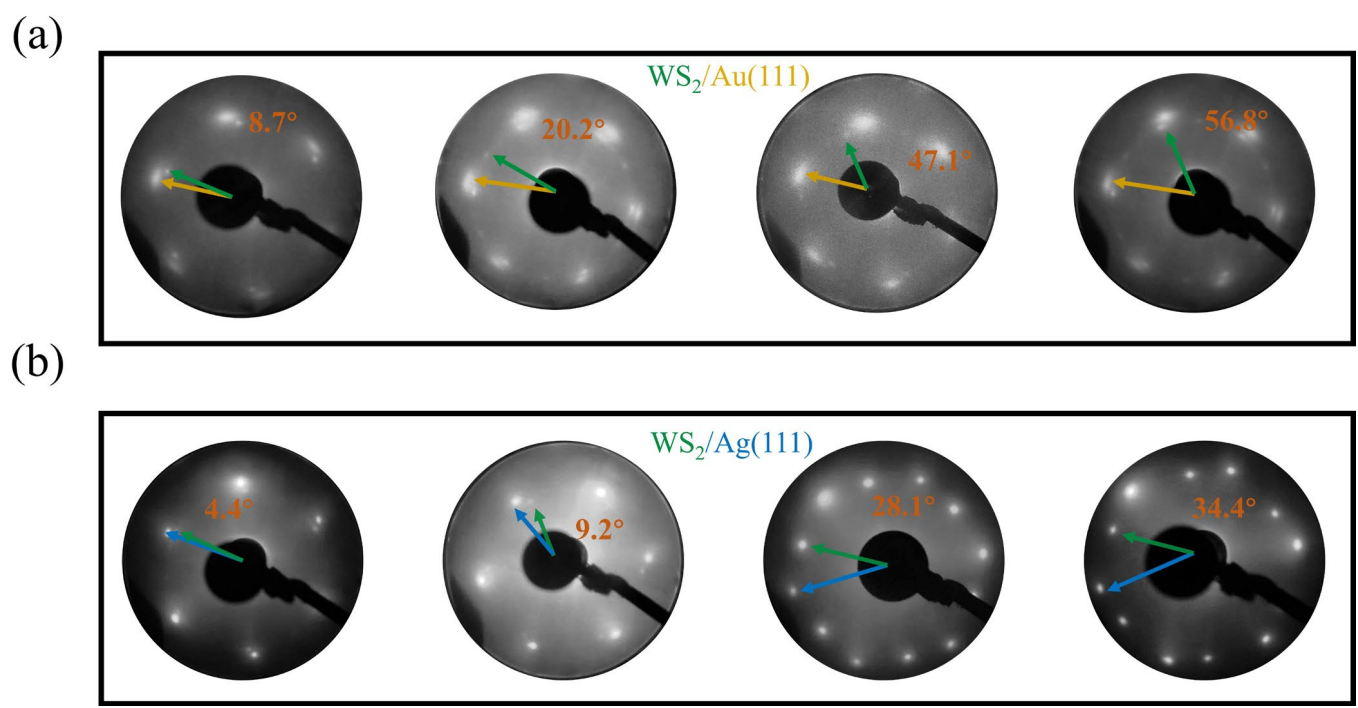

Figure S1: LEED characterization of exfoliated $WS_2$: LEED images of four different flakes of $WS_2$ on (a) Au(111) and (b) Ag(111). All images were obtained with an electron energy of 125 eV. The green arrow points to a diffraction spot of $WS_2$, while the yellow and blue arrows indicate the diffraction spots of Au and Ag, respectively. The angle between $WS_2$ and the substrate is specified for each flake.

## 2 OPTICAL MICROSCOPY

**Figure S2 (a,b)** shows representative flakes of $WS_2$ and **Figure S2(d,e)** $WSe_2$ on Au(111), along with their lateral size distribution, **Figure S2(c,f)**. The results indicate that $WSe_2$ is more easily exfoliated compared to $WS_2$.
**Figure S3** presents samples exfoliated on Ag(111), where **Figure S3(a-c)** depict three $WS_2$ samples, and **Figure S3(d-f)** depict three $WSe_2$ samples.

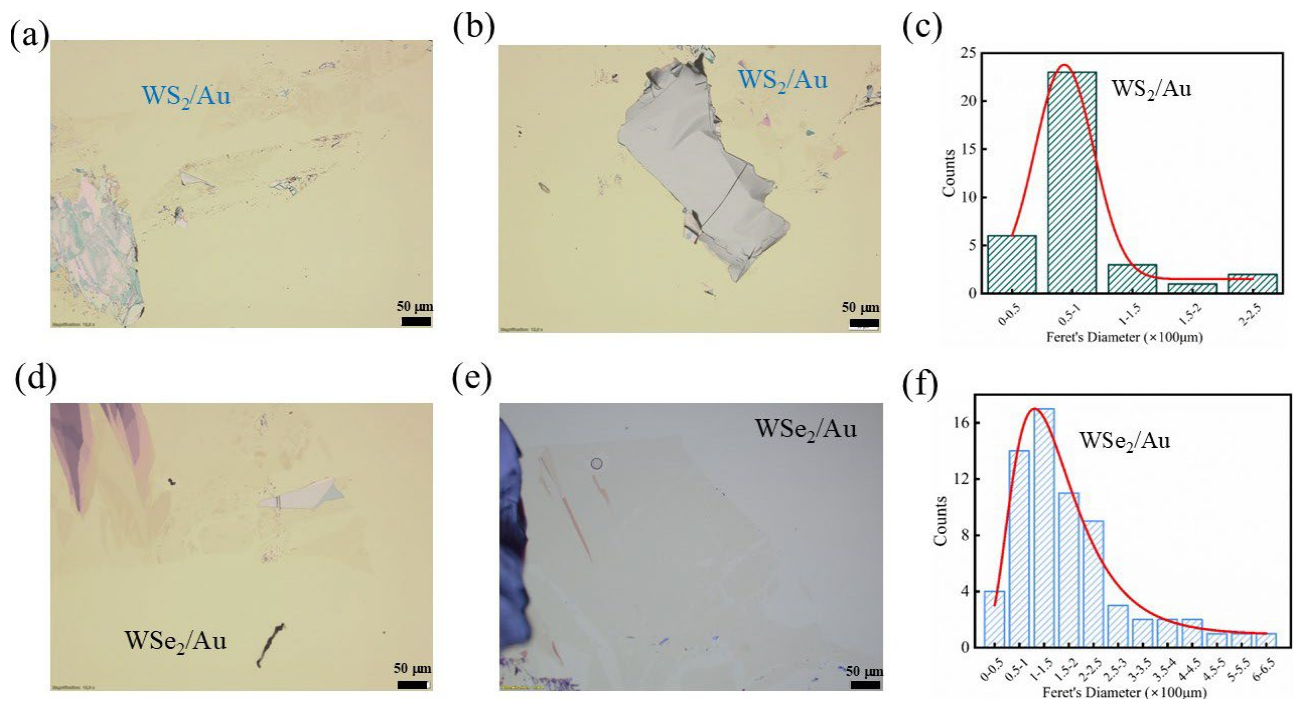

Figure S2: Optical microscopy and flake size analysis for WS$_2$ and WSe$_2$ on Au(111): (a) and (b) show two samples of WS$_2$ on Au(111), while (c) illustrates the lateral size distribution of the flakes. (d-f) are corresponding images for WSe$_2$ on Au(111).

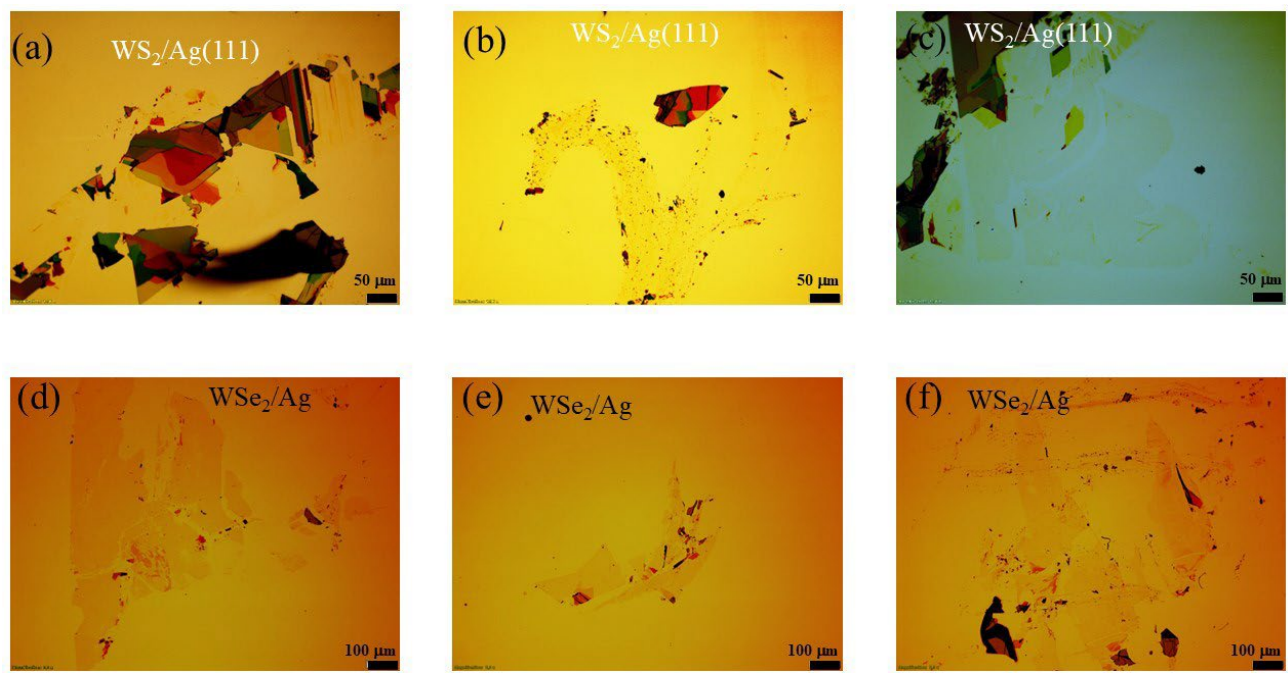

Figure S3: Optical microscopy images of WS$_2$ and WSe$_2$ on Ag(111): (a-c) show three samples of WS$_2$, and (d-f) show three samples of WSe$_2$. Color balance of the images was adjusted to make monolayers more visible.

## 3 X-RAY PHOTOELECTRON SPECTROSCOPY (XPS)

The XPS spectra of WS$_2$ on Ag(111) and Au(111) measured using a lab based spectrometer with a spot size of 1 mm$^2$ are shown in **Figure S4**. The characteristic peaks of the W 4*f* and S 2*p* orbitals confirm the presence of WS$_2$ in the sample and further indicate that the flake size meets the measurement requirements. Table S1 summarizes the binding energy values for W 4*f*, Se 3*p* and S 2*p* peaks from the XPS data presented in the main text and **Figure S4**. The data is in good agreement with values reported in the literature. Additionally, the components of oxidized W and S, expected at binding energy of 36.1 eV and
169.1 eV, respectively (3), are absent, confirming the air stability of exfoliated WS$_2$.

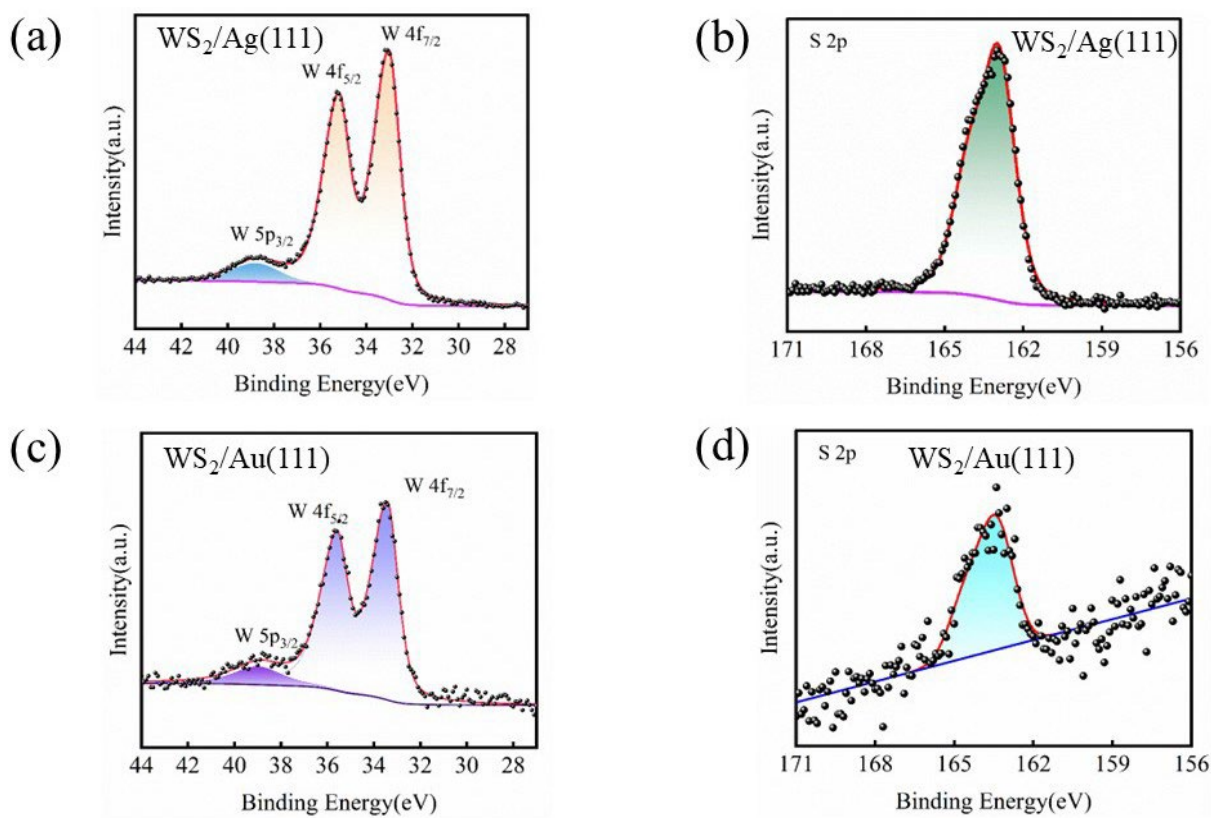

Figure S4: XPS spectra of WS$_2$ on Ag(111) and Au(111): Detailed scans of the (a) W 4*f* and (b) S 2*p* core level regions of WS$_2$/Ag(111). (c-d) detailed scans of the (c) W 4*f* and (d) S 2*p* core level regions of WS$_2$/Au(111). No signs of oxidation are visible for either W or S.

## 4 X-RAY DIFFRACTION

The single crystal X-ray diffraction measurements were performed at the I09 beamline at the Diamond Light Source (4) using the UHV end-station designed for hard X-ray photoelectron spectroscopy and X-ray standing wave studies. The beam size was about 40 (H) x 20 (V) $\mu$m at the sample. After undergoing KISS exfoliation, the Ag substrate was aligned to focus the X-ray beam to a WSe$_2$ flake by monitoring the W 4*f* photoelectron intensity. The Ag(111) Bragg reflection was subsequently excited in a near-backscattering geometry by a photon beam of 2627 eV, which was delivered by an undulator and a double-crystal Si(111) monochromator. The reflected beam was visualized by a fluorescent screen and the intensity of the beam spot was recorded by a CCD camera. **Figure S5 (d)** shows the measured Ag(111) reflectivity curve and the best fit based on dynamical theory of X-ray diffraction taking into account the contribution of the beamline optics. The FWHM of the peak was determined to be 0.95 eV, which is identical to the intrinsic width of the (111) reflection of a perfect Ag single crystal convoluted with the energy width of the Si(111) monochromator. This confirms our exfoliation process does not introduced deformation to the substrate lattice.

|  | WSe$_2$/Au(111) (eV) | WS$_2$/Au(111) (eV) | WS$_2$/Ag(111) (eV) |
|---|---|---|---|
| W 4$f_{7/2}$ | 32.50 | 33.42 | 33.11 |
| W 5$p_{3/2}$ | 38.01 | 38.97 | 38.48 |
| Se 3$p_{3/2}$ | 161.21 | / | / |
| S 2$p_{3/2}$ | / | 163.38 | 162.91 |

**Table S1**. Binding energies of the core level photoemission lines for W, S and Se extracted from the spectra in Figure 5 and Figure S4. The XPS spectra were collected under the same conditions.

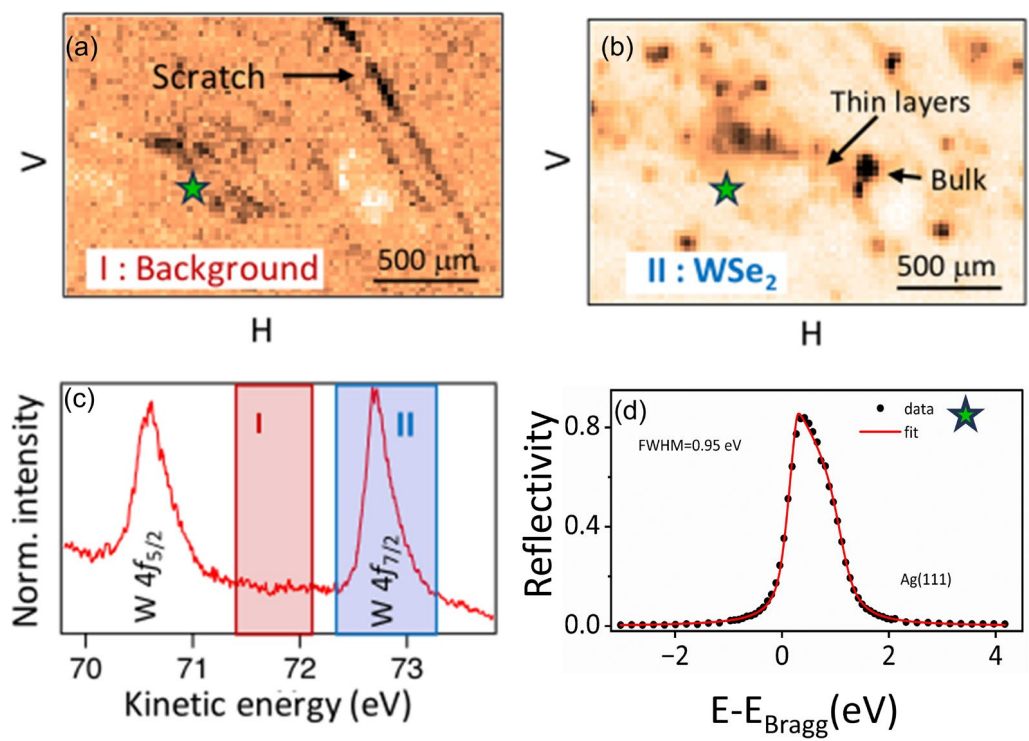

Figure S5: X-ray diffraction of WSe$_2$ on Ag(111): (a-b) Presents images of the background (region I) and WSe$_2$ flakes (region II). (c) photoemission spectrum of the W 4$f$ core level region used to select the spots marked in (a) and (b) with green stars. (d) The rocking curve of Ag(111), with FWHM = 0.95 eV.

## 5 BULK CRYSTAL QUALITY

The quality of the bulk crystal may also influence the success of the KISS exfoliation. Figure S6 presents the WSe$_2$ and WS$_2$ crystals used for KISS exfoliation in this work. The WSe$_2$ crystal, **Figure S6(a)**, consists of two large, flat, and smooth terraces, whereas the WS$_2$ crystal, **Figure S6(b)**, features smaller grains within the larger terraces.

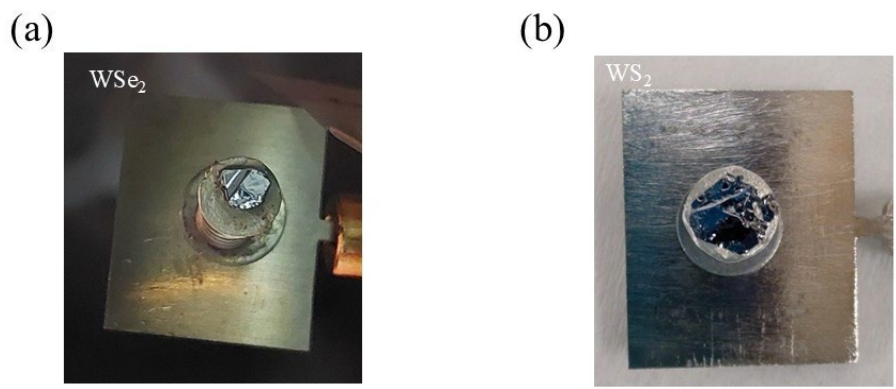

Figure S6: Images of bulk WS$_2$ and WSe$_2$ crystals: (a) shows the bulk WSe$_2$ crystal with large, flat and smooth surface. (b) shows the surface of WS$_2$ bulk crystal that appears of lower quality with several smaller grains embedded in large, flat terraces.